\documentstyle[12pt,epsf]{article}
\begin{document}
\begin{titlepage}
\pagestyle{empty}
\baselineskip=21pt
\rightline{Alberta Thy-1-94}
\rightline{August 1994}
\vskip .2in
\begin{center}
{\large{\bf  Biased Discrete Symmetry Breaking and Fermi Balls}}
\end{center}
\vskip .1in
\begin{center}
Alick L. Macpherson and Bruce A. Campbell

{\it Department of Physics, University of Alberta}

{\it  Edmonton, Alberta, Canada T6G 2J1}

\vskip .2in

\end{center}
\centerline{ {\bf Abstract} }
The spontaneous breaking of an approximate discrete symmetry is considered,
with the resulting protodomains of true and false vacuum being separated by
domain walls. Given a strong, symmetric Yukawa coupling of the real scalar
field to a generic fermion, the  domain walls accumulate a gas of  fermions,
which   modify the domain wall dynamics.  The splitting of the degeneracy of
the ground states results  in the false vacuum protodomain structures
eventually being fragmented into tiny  false vacuum bags with a Fermi gas shell
(Fermi balls), that may be cosmologically stable due to the Fermi gas pressure
and wall curvature forces, acting on the domain walls.  As fermions inhabiting
the  domain walls do not undergo number density freeze out, stable Fermi balls
exist only if  a fermion anti-fermion asymmetry occurs. Fermi balls formed with
a new Dirac fermion  that possesses no standard model gauge charges provide a
novel cold dark matter candidate.
\baselineskip=18pt
%%%%%%%%%%%%%%%%%%%%%%%%%%%%%%%%%%%%%%%%%%%%%%%%%%%%%%%%%%%%%%%%%%%%%
%%%%%%%%%%
\end{titlepage}
%\newpage
\baselineskip=18pt

It is well known that spontaneous breaking of a discrete symmetry can produce
topological structures composed of different domains separated by topological
defects \cite{1,2,3}. In the simplest  such physical scenario, the
topological defects produced are domain walls \cite{1}  (transition regions
between spatial domains that possess topologically different vacuum
orientations), which within the context of  cosmological models,  have been
applied to phenomena ranging from  energetically soft topological defects
\cite{4} and structurons \cite{5,6}  for  the formation of large scale
structure \cite{7,8}, to significant deviations from thermal equilibrium at the
QCD scale \cite{9},  neutrino balls \cite{10,11}, and an origin for
cosmological Gamma Ray Bursts \cite{12}. In this paper, the interaction of
domain walls with a fermion sector is considered, which suggests the possible
production of composite microscopic cosmological relics referred to henceforth
as Fermi balls.  These Fermi balls, under certain  conditions, provide an
unusual  source for cold dark matter, and may be relics of the seeds for
possible structure formation in the cold dark matter scenario.

The  simplest model  exhibiting topological structure is that of a real scalar
field $\varphi$ with a Lagrange density of the form
\begin{equation}
{\cal L} = \frac{1}{2} \partial_{\mu} \varphi \partial^{\mu} \varphi -
           \frac{\lambda^{2}}{8}(\varphi^{2}-{\varphi}_{0}^{2})^{2}
\label{lag1}
\end{equation}
Clearly, equation (\ref{lag1}) possess a $Z_{2}$ symmetry (invariance under
$\varphi \rightarrow - \varphi$), which if spontaneously broken results in a
vacuum  expectation value (VEV) for $\varphi$ that has two possible values;
$<\varphi> = \pm {\varphi}_{0}$. These  two VEV's  correspond to topologically
distinct  vacuum orientations (distinct values of the order parameter); here
the notion of   topologically distinct vacua implies that  one vacuum
orientation cannot be continuously  deformed into the other. Yet due to the
$Z_{2}$ symmetry being exact,  neither VEV is  preferred, so the determination
of the VEV  in a particular spatial region is set by random fluctuations in
$ \varphi$.  Thus the spontaneous symmetry breaking results in  a randomly
generated   network  of spatial domains of both  vacuum orientations that are
separated by transition regions called  domain walls ( topological defects).
The form of the domain wall solution is a topological soliton of class
${\pi}_{0}$ \cite{13}, and is easily obtained from  the equation of motion for
$\varphi$. The simplest  such solution is that of a planar domain wall in the
xy plane at $z=0$ with the boundary conditions $\varphi( z \rightarrow \pm
\infty)= \pm {\varphi}_{0}$, and has the form  $<\varphi> =  {\varphi}_{0}
\tanh(\frac{z}{\delta})$. Here $\delta = \frac{2}{\lambda {\varphi}_{0}}$ is
the wall thickness. Typically,  $\delta$ is  assumed to be small compared to
the  average radius of curvature of the walls (the thin wall  approximation ),
so that the   domain walls can be treated as two dimensional surfaces. For the
planar  domain wall, the associated stress-energy tensor  is $T^{\mu}_{\nu}=
{{\lambda}^{2} {{\varphi}_{0}}^{4} \over 4}\cosh^{ -4}(\frac{z}{\delta})$
diag($1,1,1,0$), indicating that the only  non zero pressure components are
within the plane of the wall, and both are equal to minus the energy density.
Due to the form of the stress energy tensor, the surface tension
($ \int{T^{i}_{i}dz} $) is  exactly equal to the surface energy density of the
wall ($\int {T^{0}_{0} dz}$), which has the form
\begin{equation}
\sigma={{2 \lambda {\varphi}_{0}^{3}} \over 3}
\label{surfendens}
\end{equation}

The  cosmological implication of  spontaneous breaking of an exact discrete
symmetry, as first   analysed by Zel'dovich et al. \cite{1}, is the formation
of stable domain walls separating  protodomains (spatial regions  with distinct
vacuum orientations) of topologically distinct energetically degenerate ground
states. These walls  evolve to planar structures that dominate the energy
density of the Universe. Clearly, this is in  contradiction with our present
observations.  To avoid this prediction, the self coupling of $ \varphi$ could
be fine tuned so to sufficiently delay the wall dominance of the energy
density. A more creditable alternative, suggested in \cite{1} is to remove
the  wall stability by  requiring the discrete symmetry to be only approximate.
Then, spontaneous symmetry breaking results  in topologically distinct  ground
states that are non degenerate, as the symmetry breaking is biased. This non
degeneracy manifests itself in the form of  protodomains of true and false
vacuum, that are separated by domain walls.

Upon formation, the domain walls evolve  in accordance with the protodomain
ensemble minimising its energy, so that the wall motion can be described  in
terms  of the pressure imbalance across the domain wall \cite{14}.  In a
$ \varphi$ self coupling model \cite{14}, only the false vacuum volume
pressure and the normal component of the wall surface tension contribute  to
the pressure imbalance. The  false vacuum volume pressure is typically
constant, and pulls   the wall towards the      false vacuum protodomain,
whilst the normal component of the surface tension acts to straighten the wall,
and decreases with decreasing wall curvature. Thus,  finite sized false vacuum
protodomains (vacuum bags) collapse on themselves, whilst infinite domain walls
are pulled  toward the false vacuum region \cite{14}.  It is this  biased
discrete symmetry breaking, with its inevitable conversion of false  to true
vacuum that  cause domain walls to disappear, and by which  a wall dominated
energy density disaster is avoided \cite{2,15,16}. Obviously,
the degree of biasing between the vacua dictates the average domain wall
lifetime, and if their longevity is sufficient for them to dominate the energy
density of the Universe, then power law inflation can be induced
\cite{2,1,15}.

As no $ \varphi$ self coupling model of biased discrete symmetry breaking
produces  stabilised finite size vacuum bags from topological defects, other
more novel couplings have been investigated \cite{17,10,18}, each with
there own cosmological implications. The coupling advocated in this paper is
one which relies on the presence of fermions strongly coupled to the scalar,
with $\varphi$ symmetrically coupled to a fermion via standard  Yukawa
couplings:
 \begin{equation}\label{lag3}
{\cal L} = \frac{1}{2} \bar{\psi}(i \partial - G \varphi ) \psi  +
\frac{1}{2} \partial_{\mu} \varphi \partial^{\mu} \varphi -
 \frac{\lambda^{2}}{8}(\varphi^{2}-{\varphi}_{0}^{2})^{2} +A(\varphi)
\end{equation}
The lagrange density now contains both a Yukawa coupling of fermions to the
scalar field $\varphi$, and  a term $A(\varphi)$ that explicitly breaks  the
discrete symmetry to an approximate one. The actual form of $A(\varphi)$ is
specified only to the extent that the energy difference between the two VEV
orientations is $\Lambda $. ( For specific examples of $A(\varphi)$ consult
\cite{17} .) The  Yukawa coupling implies that after spontaneous
breaking,  fermions acquire a mass proportional to  $ <\varphi>$, and so it is
energetically favourable for the fermions to inhabit the domain wall as they
become effectively massless there. (In the infinite planar wall there exists an
analytic  solution for the zero mode of the fermion bound to the domain wall
\cite{19}.) Thus, any off wall fermions  (that are strongly coupled) are
swept up by, and reside in  the domain wall. Since immediately after the phase
transition  each fermion will be, on average, within a correlation length of the
percolating wall structure, we expect the fermions to be efficiently stuck to
the walls.  Domain walls quickly become populated with fermions, so that the
walls (in the thin wall approximation) are essentially   two dimensional
surfaces inhabited by a Fermi gas of massless fermions. The associated  Fermi
gas pressure contributes to the pressure imbalance  and acts to modify the wall
dynamics. In order to halt the collapse of a finite sized false vacuum
protodomain, and  give stable false vacuum bubbles, the Fermi gas pressure must
cancel the   surface tension and false vacuum  volume pressure components.
This will occur if the energy of  a false vacuum protodomain that has
accumulated  a wall gas of $ N$ fermions can be  minimised for some  finite
radius.  For a vacuum bag of arbitrary shape, the energy of the bag is
\begin{equation}\label{En1}
E= V \Lambda + S \sigma +{E}_{F}
\end{equation}
($V= $ the volume of the vacuum bag, $S=$ its surface area, and ${E}_{F}=$ the
energy of the Fermi gas composed of $N$ wall fermions.) Assuming the wall gas
is   composed of massless degenerate fermions with g=2 internal degrees of
freedom, ${E}_{F}$ in the zero temperature limit is
\begin{equation}\label{EF}
{E}_{F}={{4 \sqrt{\pi}  {N}^{{3 \over 2}}}\over 3 \sqrt{g} \sqrt{S}  }
\end{equation}
 and for   a spherical vacuum bag,   a  stabilised bag is found, with a radius
given by
\begin{equation}\label{minR}
{N}^{{3 \over 2}} =  6 \pi \sqrt{g} \left( {R}^{4} \Lambda + 2 {R}^{3} \sigma
\right)
\end{equation}
This halting of the collapse process is due entirely to  the presence of the
fermions on the domain wall, and so for  spherical false vacuum protodomains,
one might expect stabilised    false vacuum bags with a bounding outer skin of
massless fermions.

But  assuming the  collapse of  false vacuum bags to be  completely described
by the process of spherical shrinking until the pressure imbalance is nullified
is incorrect. The collapse process is driven by a minimisation of the bag
energy, to which  there are  three competing elements: volume energy density
splitting, surface tension energy, and surface Fermi gas energy. As the surface
tension energy and the energy of the  two dimensional Fermi gas, $ {E}_{F}$,
are  dependent on  the surface area  of the bag and not its volume  (equation
(\ref{EF})), the vacuum bag energy can be reduced by  a decrease in  the bag
volume, with the surface area  held constant.  Thus, bags are unstable with
respect to ``pancake'' deformations, implying  the  bag flattens into a
sheet-like structure. In conjunction with this flattening,  the  vacuum bag lowers
its energy by fragmenting into smaller vacuum bags. To see that fragmenting is
favoured, consider the energy for an arbitrary vacuum bag, but first  neglect
the volume contribution. By minimising  this energy with respect to the surface
area $S$,  the energy of the stabilised bag is found to be
\begin{equation}\label{EminS}
{E \mid}_{  V\Lambda =0}= 3\left( {4 \sigma \pi  \over  9 g } \right)^{{1 \over
3}} N
\end{equation}
which is proportional to $ N$. This implies that  one vacuum bag with a domain
wall Fermi gas composed of  $N$ fermions  is energetically equivalent to two
vacuum bags each with ${N \over 2}$ fermions on their  domain wall, and so
vacuum bags may  fragment but are not compelled to do so.  However, on
inclusion of the false vacuum volume energy, minimisation of energy favours bag
fragmentation.  These facets of the collapse process for a finite sized false
vacuum protodomain result in  a more involved vacuum bag evolution than the
simple shrinkage to a minimal surface area stabilised by $ N$ wall inhabiting
fermions, as all three act concurrently.  The physical collapse process of a
false vacuum bag is one of repeated shrinking, flattening, and fragmenting,
that results in numerous smaller vacuum bags.

 However, for a sufficiently strong coupling of the fermions  to the scalar order parameter
the collapse process does not continue  ad infinitum, as the soliton origin of
the bag structure will  eventually arrest the collapse. This onset of the
quantum regime is signified by the breakdown  of the thin wall approximation,
and implies  that the domain wall radius of curvature is comparable to the size
of the vacuum bag.  When this occurs, the vacuum bag is no longer a bubble of
false vacuum with a domain wall skin containing a two dimensional Fermi gas,
but rather  a ball composed almost exclusively of the domain wall, with almost
all the interior false vacuum  having been destroyed. Such a ball of domain
wall still carries the Fermi gas, but now the massless fermions of the  Fermi
gas constitute a three dimensional Fermi gas inhabiting the interior of the
domain wall ball.  It is these balls of  fermion populated domain wall that we
refer to as Fermi balls, and they represent  true non topological defects. If
the fermions are strongly coupled to the scalar, then the Fermi balls will be
stable if the energy invested in the scalar field configuration is less than
the total mass the trapped fermions would have to obtain if the wall
disappeared. To get a crude estimate of  the size of the stabilised Fermi balls,
we note that our Lagrangian contains only one dimensional parameter which, in
the wall solution, determines its intrinsic thickness. By equating the minimum
size of the  stabilised Fermi balls $ {R}_{min} $ to the wall thickness
$\delta$, and assuming these stabilised Fermi balls adopt a minimum surface
area configuration,  the typical stabilised radius (radius at which the
collapse process stops) is estimated by
\begin{equation}\label{Rmin}
{R}_{min} \sim \frac{2}{\lambda {\varphi}_{0}}
\end{equation}
The radius  $ {R}_{min} $ is small, indicating the collapse of false vacuum
protodomains produces in a mist of tiny Fermi balls distributed  throughout the
3-space. This mist of Fermi balls should be  considered as possible
cosmological relics, since their  stability against further collapse  may be
assured by energetic considerations, and Fermi ball annihilation is ruled out
if the fermions are  Dirac particles with conserved fermion number.

Yet biased spontaneous  symmetry breaking doesn't necessarily result in the
formation  of finite sized false vacuum protodomains. The nature of the
protodomain structure at formation depends on the degree of anomalous breaking
$ \Lambda$;  for $ \Lambda$ small  compared to the potential barrier,  a
percolating  domain wall structure \cite{20,3,8} is expected, whilst a $
\Lambda$  comparable to the barrier height implies  the formation of  finite
sized false vacuum bags.  For the dynamical evolution of percolating domain
walls, the analysis and conclusions differ little from that  of  Gelmini et.
al. \cite{14}, who show that  although there are several different
cosmological scenarios, in which the domain walls straighten out on various
scales, the false vacuum volume pressure eventually dominates the pressure
imbalance. This causes  the domain walls to be driven inward on the false
vacuum protodomain structure,  inducing  a ``melting'' of the false vacuum.
Once the false vacuum volume pressure becomes dominant, the conversion of false
to true vacuum is relentless, and eventually leads to a fragmentation of the
percolating domain wall structure into finite sized false vacuum bags. This
fragmentation is  essentially the  conversion of topological defects to
nontopological ones, and  is a result of  the system's desire to
minimise its energy.  Inclusion of a  strong coupling to a fermion sector
causes a modification to the constraints on  $ \Lambda$ that define the
different dynamical regimes (The surface tension $ \sigma$ is replaced by  $
\sigma -  {\cal P}$ to account for the two dimensional Fermi gas pressure
${\cal P}$.), but the conclusions  of \cite{14} remain unaltered. This
implies that irrespective of the protodomain structure formed at symmetry
breaking,  finite  sized false vacuum bags are eventually produced, which in
turn evolve into the Fermi ball structures discussed above.

Thus, biased discrete symmetry breaking with strongly coupled Dirac fermions
may result in a mist of nontopological objects (Fermi balls) comprised of  a
superposition of the massless fermions and a local deformation of the order
parameter $ <\varphi >$. These Fermi balls are expected to be approximately
spherical, with a radius $ {R}_{min}$ (equation (\ref{Rmin})). Their stability
against further collapse is  assured by sufficiently strong spinor scalar
coupling, but  stability under fermion anti-fermion annihilation has not been
addressed. Such annihilations could significantly affect the Fermi ball
lifetime.

 A strong Yukawa coupling implies that after the symmetry breaking, the
fermions collect  on the domain walls, thereby enhancing the  fermion
anti-fermion annihilation rate. Fermion anti-fermion  annihilations reduce the
Fermi gas pressure, so destabilising the false vacuum bag so that collapse
continues until the pressure balance is restored. Thus, confinement of the
fermions to the domain wall prohibits freeze out of the number density of
fermions, and so Fermi balls can exist only if there is a net fermion
anti-fermion asymmetry.

Given that Fermi balls are produced, equations (\ref{minR})  and (\ref{Rmin})
imply that they would be composed of  approximately $  50$  fermions,
independent of the symmetry breaking scale,   and possess a mass of  the order
of $100 {\varphi}_{0}$ GeV. This suggests that a Fermi ball would appear as  a
very heavy  slow moving  particle, which if the individual wall  fermion had
electric charge,  would carry a  charge in the order of $10 - 50$ times the
electron charge. Such objects therefore have characteristics similar to either heavy  ions or nuclearites
\cite{21}, and so analysis of the Fermi ball  stopping power \cite{22} and the negative results of nuclearite  searches by
collaborations such as  MACRO \cite{23} can place  a constraint on the
relation between the Fermi ball mass and number density.

Alternatively, Fermi balls could be  composed of a new Dirac fermion that
possesses no  standard model gauge charges.  Fermi balls would then be neutral,
heavy, and non relativistic, and due to their  absence of gauge charges, would
interact  extremely weakly with standard model matter; barring new couplings,
the only interaction (apart from the gravitational one) would be via  couplings
of the real scalar field $ \varphi$ to the standard Higgs fields. Thus, the
heavy non relativistic neutral Fermi balls would constitute an ideal candidate
for cold dark matter. This suggests a possible constraint on these neutral Fermi
balls, as gravity  results in an accumulation of Fermi balls around massive
objects such as the sun.  Gravitationally bound Fermi balls may orbit  through
or within the solar interior, thereby transporting energy away from the solar
core by their weak scattering from solar core baryons (protons). If such heat
diffusion is sufficiently efficient, the gravitationally bound Fermi balls
become  incompatible with the standard solar model.

\begin{figure}[t]
\epsffile[-100 0 220 204]{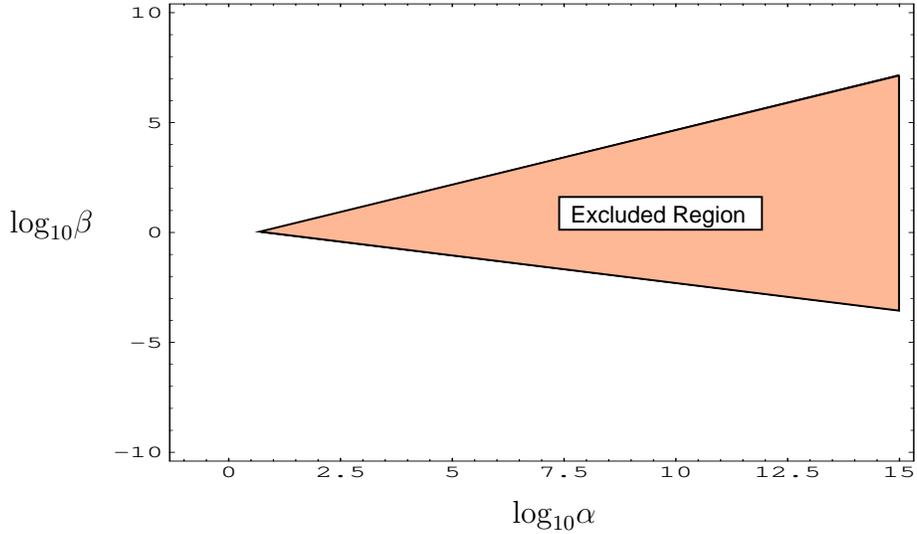}
\begin{picture}(0,0)(0,0)
\thinlines
\put(70,120){$ {\log}_{10} \beta$}
\put(260,10){$ {\log}_{10} \alpha$}
\end{picture}
\caption{ A contour plot of the relative solar luminosity carried by the Fermi
balls, as a function of $ \alpha={{m}_{FB} \over {m}_{p}}$, the Fermi ball mass
relative to the proton mass, and $ \beta= {{\sigma}_{FB} \over {\sigma}_{c}}$,
the Fermi ball-baryon  cross section, relative to the fiducial cross section $
{\sigma}_{c}\equiv{{m}_{p} \over {M}_{\odot}} {R}_{\odot} = 4.0 \times
{10}^{-36} {\mbox{cm}}^{2}$. The contour shown is that of relative luminosity
of unity, and the excluded region is where the relative luminosity is greater
than 1.\label{fig1}}
\end{figure}

Assuming the Fermi balls are  the sole source of dark matter and that their
contribution is such  that the Universe attains closure density ($ \Omega =1$),
the magnitude of the luminosity diffusion, as a function of the Fermi ball
mass, can be evaluated.  The analysis is based on the work of  Press and
Spregel \cite{24,25},  which deals with the solar  capture and the
subsequent luminosity transport of  cosmions \cite{26}.   For this
closure density scenario, with  gravitationally bound Fermi balls in
approximate thermal equilibrium with the solar core, Figure \ref{fig1} shows a
contour plot of the luminosity transported by them relative to  the solar
luminosity, as a function of the Fermi ball mass relative to the proton mass,
and the  Fermi ball-baryon cross section relative to a fiducial cross section
of reference  \cite{25}. A relative luminosity contour of unity is used to
restrict the relative cross section and relative mass of the Fermi balls (which
in turn can be related to  $ {\varphi}_{0}$), as a relative luminosity of
unity or greater implies that for fixed total energy transport, the Fermi ball
transport would more than halve the core temperature gradient, in contradiction
with the solar model \cite{25}. The restriction on parameter space isn't
particularly severe, considering that neutral Fermi balls are expected to have
extremely weak non-gravitational  interactions, and so would free stream
through the sun.

Finally, if Fermi ball closure density  is assumed, the fermion anti-fermion
asymmetry required just prior to the biased spontaneous symmetry breaking in
order to produce  Fermi balls can be estimated.  The constraint of closure
density sets a restriction on the present day Fermi ball number density, which
is  in turn related to the relative fermion anti-fermion asymmetry just prior to
symmetry breaking, defined by
\begin{equation}\label{relass}
B={{n} - {\bar{n}} \over {n}}~~~~~~~~~~~
\end{equation}
(here $ {n}$ and $ {\bar{n}}$ represent the number density of fermions and
anti-fermions). The present day Fermi ball number density is obtained from the
number density of excess fermions at the symmetry breaking by evolving this
number density forward to the present day,   and  then dividing this number
density by the number of fermions in a typical Fermi ball. From this, the
constraint on the relative fermion asymmetry is  found to be of order
\begin{equation}\label{assy}
B \sim {{10}^{-7} \mbox{GeV} \over {\varphi}_{0}}
\end{equation}
which for a breaking scale of $ {\varphi}_{0} = 1$GeV implies an asymmetry of
$ {10}^{-7}$,  which is of similar magnitude to the baryon asymmetry at the $
1$GeV scale.

\noindent{ {\bf Acknowledgements} } \\
\noindent  This work was supported in part by the
Natural Sciences and Engineering Research Council of Canada.


\begin{thebibliography}{99}
\bibitem{1} Ya. Zel'dovich, I. Yu. Kobzarev, and L.B. Okun, Zh. Eksp. Teor.
Fiz. {\bf67} (1974) 3. [Sov.  Phys. JETP {\bf 40} (1975)1.]
\bibitem{2} T.W.B. Kibble, J. Phys.  {\bf A9} (1976)1387.
\bibitem{3} for a review see: A. Vilenkin, Phys. Rep. {\bf 121} (1985) 263.
\bibitem{4} I. Wasserman, Phys.  Rev. Lett.  {\bf 57} (1986)  2234; C.
Hill, D. Schramm, and J. Fry, Comm on Nucl. and Part. Phys.  {\bf 19} (1989)
25; C. Hill, D. Schramm, and D. Widrow, Fermilab-PUB-89/166,  (1989).
\bibitem{5} Z. Lalak and B.A. Ovrut, CERN preprint CERN -TH 6957/03  {\bf 71}
(1993)  951.
\bibitem{6} Z. Lalak and B.A. Ovrut, Phys.  Rev. Lett.  {\bf 71} (1993)  951.
\bibitem{7}  S. Lola and G.G Ross, Nucl. Phys.    {\bf B406} (1993) 452.
\bibitem{8}  Z. Lalak, S. Lola , B.A. Ovrut, and G.G Ross, Oxford University preprint OUTO-93-22P // hep-ph 9404218.
\bibitem{9} J Preskill, S. P. Trivedi, F Wilczek, and M. B. Wise, Nucl. Phys.
{\bf B363} (1991) 207.
\bibitem{10}  B. Holdom, Phys.  Rev.   {\bf D36} (1987) 1000.
\bibitem{11}  A.D. Dolgov and O. Yu. Markin,  Sov. Phys. JETP   {\bf 71} (1990)
207.
\bibitem{12}  B. Holdom and R. A. Malaney, CITA preprint 22/93 // astro-ph 9306014.
\bibitem{13} S-T. Hu, {\it Homotopy Theory},  Academic Press, New York
(1959).
\bibitem{14} G.B. Gelmini, M. Gleiser, and E.W. Kolb, Phys. Rev. {\bf
D39} (1989)1558.
\bibitem{15}  A . Vilenkin, Phys.  Rev.   {\bf D23} (1981) 852.
\bibitem{16} P. Sikivie, Phys.  Rev. Lett.  {\bf 48} (1982) 1156.
\bibitem{17} J. Frieman, G.B. Gelmini, M. Gleiser, and E.W. Kolb, Phys.
Rev. Lett.{\bf 60} (1988) 2101.
\bibitem{18}  R. Friedberg, T. D. Lee, and  A. Sirlin,  Phys.  Rev.  {\bf
D13} (1976) 2739, and Nucl. Phys.    {\bf B115} (1976) 1, 32.
\bibitem{19} W.A. Bardeen et al.,  Phys. Rev. {\bf D11} (1975) 1094.
\bibitem{20} D. Stauffer, Phys. Rep. {\bf 54} (1979)1.
\bibitem{21} A. De R\'{u}jula and S. L. Glashow, Nature   {\bf 312} (1984)
734.
\bibitem{22} A.L. Macpherson and J.L. Pinfold, in preparation.
\bibitem{23} MACRO Collaboration, Phys.  Rev. Lett.  {\bf 69} (1992) 1860.
\bibitem{24} W.H. Press and D.N. Spregel, Astrophys.  J.  {\bf 296} (1985)
679.
\bibitem{25}  D.N. Spregel and W.H. Press, Astrophys.  J.  {\bf 294}
(1985) 663.
\bibitem{26} G. Steigman, C.L. Sarazin, H. Quintana, and J. Faulkner,
Astrophys.  J.  {\bf 83} (1978) 1050.

\end{thebibliography}
\end{document}